\documentclass[fleqn,10pt]{wlscirep}

\title{Electron--phonon Coupling and the Superconducting Phase Diagram of the LaAlO$_3$--SrTiO$_3$ Interface}
\author[1,*]{Hans Boschker}
\author[1,2]{Christoph Richter}
\author[1]{Evangelos Fillis-Tsirakis}
\author[3]{Christof W. Schneider}
\author[1]{Jochen Mannhart}

\affil[1]{Max Planck Institute for Solid State Research, 70569 Stuttgart, Germany}
\affil[2]{Center for Electronic Correlations and Magnetism, Augsburg University, 86135 Augsburg, Germany}
\affil[3]{Paul Scherrer Institute, 5232 Villigen, Switzerland}

\affil[*]{h.boschker@fkf.mpg.de}

\begin{abstract}

The superconductor at the LaAlO$_3$--SrTiO$_3$ interface provides a model system for the study of two-dimensional superconductivity in the dilute carrier density limit. Here we experimentally address the pairing mechanism in this superconductor. We extract the electron--phonon spectral function from tunneling spectra and conclude, without ruling out contributions of further pairing channels, that electron--phonon mediated pairing is strong enough to account for the superconducting critical temperatures. Furthermore, we discuss the electron--phonon coupling in relation to the superconducting phase diagram. The electron--phonon spectral function is independent of the carrier density, except for a small part of the phase diagram in the underdoped region. The tunneling measurements reveal that the increase of the chemical potential with increasing carrier density levels off and is zero in the overdoped region of the phase diagram. This indicates that the additionally induced carriers do not populate the band that hosts the superconducting state and that the superconducting order parameter therefore is weakened by the presence of charge carriers in another band.  
 
\end{abstract}
\begin{document}
\flushbottom

\maketitle
\thispagestyle{empty}

\section*{Introduction}

Interface superconductors are model systems for superconductivity in reduced dimensions \cite{Pereiro2011} and provide input to the long-standing problem of understanding the mechanism of superconductivity in the layered cuprate superconductors. A prominent example is the superconductor at the LaAlO$_3$--SrTiO$_3$ interface \cite{Ohtomo2004, Reyren2007, Caviglia2008}. In this two-dimensional electron liquid (2DEL), superconductivity coexists with ferromagnetism \cite{Li2011, Bert2011, Dikin2011}  and spin-orbit coupling  \cite{Caviglia2010rashba,Zhong2013}, providing the ingredients for exotic superconducting states such as finite momentum pairing \cite{Michaeli2012,Loder2013}. Regarding the pairing mechanism, both conventional electron--phonon coupling and electronic pairing mechanisms are being considered \cite{Klimin2014,Stephanos2011, Scheurer2015}. We recently mapped out the superconducting gap across the phase diagram and obtained a picture qualitatively similar to the phase diagram of the high-$T_\textrm{c}$-cuprate superconductors: in the underdoped region the gap increases with charge carrier depletion \cite{Richter2013}. This similarity between the doping dependence of the superconducting gap of LaAlO$_3$--SrTiO$_3$ and the high-$T_\textrm{c}$-cuprates makes understanding the pairing mechanism in LaAlO$_3$--SrTiO$_3$ even more relevant. No experimental study of the pairing interaction exists, however. 

The LaAlO$_3$--SrTiO$_3$ interface 2DEL differs from the electron system in doped SrTiO$_3$ because the band structures of the systems are different. But the critical temperatures are in the same range and it may well be that the pairing mechanism is the same. The superconductivity in doped SrTiO$_3$ \cite{Schooley1964} is generally explained in terms of the Bardeen Cooper Schrieffer (BCS) theory. Different phonon modes can contribute to the pairing interaction and the relative contributions of the different modes are still a topic of discussion \cite{Appel1969, Ngai1974, Takada1980, Baratoff1981, Klimin2012}. The critical temperature $T_\textrm{c}$ of doped SrTiO$_3$ has a dome-shaped dependence on the carrier density $n$, with a maximum $T_\textrm{c}$ (400-600 mK) at an extremely small carrier density of $\sim$10$^{19}$ cm$^{-3}$ \cite{Koonce1967, Pfeiffer1969, Lin2014}. According to the BCS theory, the $T_\textrm{c}$ of a superconductor depends on the Fermi surface area and on the electron--phonon coupling strength. As doped SrTiO$_3$ has a small Fermi surface area, it follows that the electron--phonon coupling would need to be exceptionally strong to explain superconductivity at such a small carrier density. Strong electron--phonon coupling is indeed possible because the Fermi energy and the plasma edge are smaller than some of the SrTiO$_3$ phonon energies (predominantly the fourth longitudinal optical mode LO4 at $\sim$100 meV) involved. Therefore plasma excitations cannot screen these phonons well, yielding strong electron--phonon coupling, as observed in tunneling experiments \cite{Sroubek1969, Hayashi1981}. At $n$ $>$ 5$\cdot$10$^{19}$ cm$^{-3}$, however, the plasma edge energy exceeds 100 meV, so that screening becomes more effective and the electron--phonon coupling is reduced. This mechanism has been proposed to explain the reduction of $T_\textrm{c}$ of doped SrTiO$_3$ in the overdoped regime \cite{Baratoff1981}. Furthermore, the SrTiO$_3$ longitudinal optical phonon modes have also been suggested to be important for the superconductivity in FeSe monolayers on SrTiO$_3$ \cite{Lee2014}.

To shed light into the pairing mechanism of the superconducting phase at the LaAlO$_3$--SrTiO$_3$ interface we have performed tunnel experiments to spectroscopically measure the electron--phonon coupling $\alpha^2F$($\omega$) at the LaAlO$_3$--SrTiO$_3$ interface. We observe coupling to the SrTiO$_3$ LO modes and conclude that these phonons likely provide the pairing interaction. We measured the evolution of both the chemical potential and the electron--phonon spectral function across the superconducting dome. The electron--phonon spectral function is found to only depend on the carrier density in the underdoped region. In the overdoped region the chemical potential is surprisingly constant. It is concluded that with increased doping the additionally induced charge carriers reside in a band that does not contribute to the pairing. Instead, the additional carriers result in Coulomb scattering of the electrons in the superconducting band, thereby reducing the superconducting gap.

\section*{Results}

\begin{figure}
\centering
\includegraphics[width=9cm]{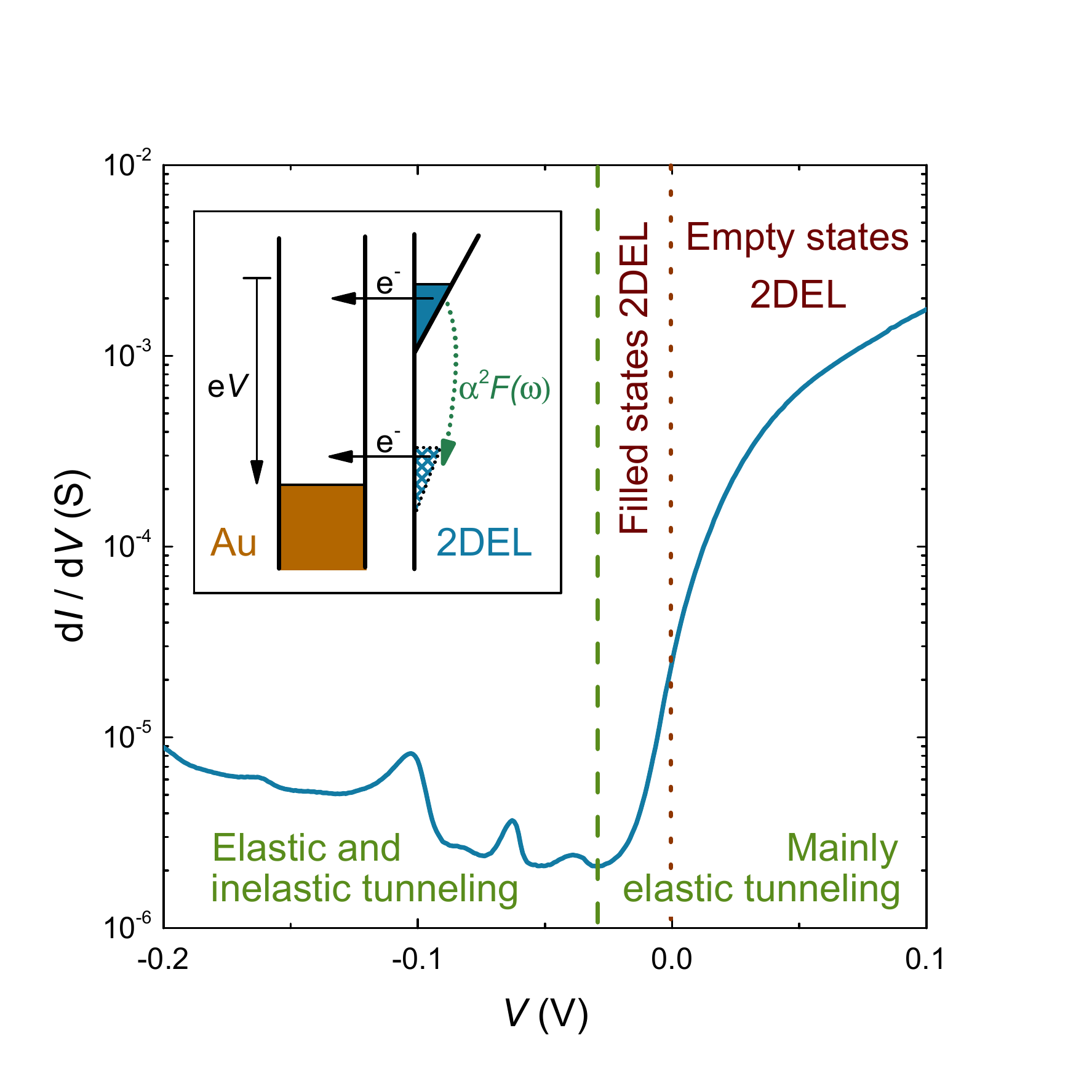}
\caption{d$I$/d$V$ ($V$) tunneling characteristic measured at $T$ = 4.2 K. The dashed green line separates the voltage ranges where elastic and inelastic tunneling dominate in the transport. For $V$ $>$ 0 electrons tunnel from the Au electrode into unoccupied states in the 2DEL and for $V$ $<$ 0 electrons tunnel from the 2DEL into the Au. In the inelastic tunneling regime, the transport is mainly by phonon--assisted tunneling with SrTiO$_3$ phonon modes. This process is illustrated in the density of states versus energy diagram of the junction in the inset, here the density of states of Au is shown on the left, that of the 2DEL on the right. If driven by voltages so large that their energy exceed phonon energies, electrons in the occupied states of the 2DEL (blue area) tunnel directly into the Au or they first make a transition to a virtual state in the gap by emitting a phonon (dashed blue area) and then transfer into the Au. Note that the linear form of the DOS($E$) shown in the schematic is not the actual DOS($E$) of the 2DEL.}
\label{fig1}
\end{figure}

Figure~\ref{fig1} presents the differential conductance characteristic $\textrm{d}I/\textrm{d}V$ ($V$) of a typical LaAlO$_3$--SrTiO$_3$ tunnel junction at $T$ = 4.2 K. Here $I$ is the tunnel current and $V$ is the voltage applied between the 2DEL and the Au counterelectrode. The polarity of the voltage characterizes the sign of the interface voltage with respect to the top electrode bias; for $V<0$ electrons tunnel out of the 2DEL. The tunnel characteristics are asymmetric, with a large tunnel conductance for $V>0$ and a relatively small tunnel conductance for $V<0$. At voltages well below the barrier height, the differential conductance of a tunnel junction is proportional to the density of states in the electrodes. Because the density of states of the Au electrode does not change significantly with energy, the $\textrm{d}I/\textrm{d}V$ ($V$) characteristic reflects the density of states of the 2DEL and the inelastic tunneling processes. Close to the conduction band edge of the 2DEL, the 2DEL density of states is strongly energy-dependent. When the absolute value of the negative voltage exceeds the chemical potential $\mu$, the differential conductance is expected to almost vanish, as the tunnel conductance then probes the density of states in the bandgap. Here $\mu$ is defined with respect to the band edge, so at $V$ = 0 V the 2DEL density of states at $E$ = $\mu$ is probed. The differential conductance rapidly decreases for $V<0$ and a minimum is observed at $V=-30$ mV (Fig.~\ref{fig1}). We identify this minimum with the conduction-band edge and attribute the increase of the conductance at larger negative bias voltages to inelastic tunneling processes and to barrier effects. In this part of the characteristic two prominent peaks are present, at $V\approx-60$ mV and $V\approx-100$  mV. 

\begin{figure}
\centering
\includegraphics[width=9cm]{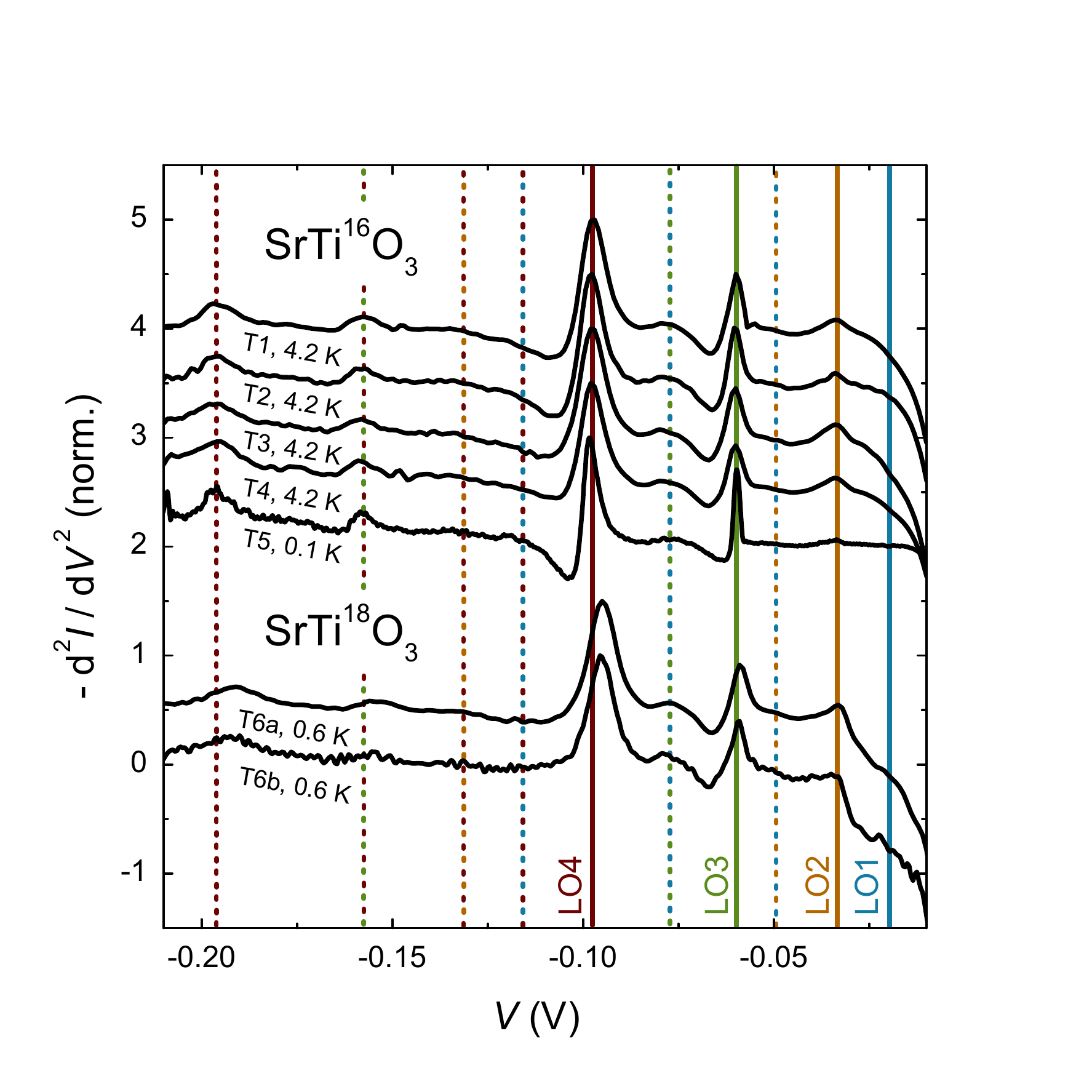}
\caption{The second derivative -d$^2I$/d$V^2$($V$) characteristics of tunnel junctions showing phonon--assisted tunneling. The data were normalized to a maximum of 1 and offset by multiples of 0.5. Samples T1, T2, T3, T4 and T6 have a 4 unit cell thick LaAlO$_3$ layer, while sample T5 has a five unit cell thick layer. The colored solid lines mark the energies of the LO phonon modes. Dashed lines mark sums and harmonics of these energies with the same color code. The two devices on LaAlO$_3$--SrTi$^{18}$O$_3$ sample are referred to by T6a and T6b. }
\label{18O}
\end{figure}

To analyze the origin of the peaks in the inelastic tunneling conductance, the second derivative of the $I$($V$) characteristics is analyzed, as the peaks in the second derivative correspond to the energies of the interacting boson modes \cite{Wolfbook}. Figure~\ref{18O} presents the -d$^2$$I$/d$V^2$($V$) characteristics of seven tunnel junctions on different samples. Five of the samples were fabricated with standard LaAlO$_3$--SrTiO$_3$ interfaces, while sample T6 was fabricated using a SrTi$^{18}$O$_3$ substrate. The five standard samples have very similar characteristics, the peaks being at identical voltages. The four LO phonon modes of SrTiO$_3$ are at energies of 19.8, 33.0, 58.8 and 98.6 meV (measured at room temperature) \cite{Spitzer1962, Cowley1964, Frederikse1967, Servoin1980, Vogt1988}. At the energies of the LO2, LO3 and LO4 modes we indeed observe peaks in the -d$^2$$I$/d$V^2$($V$) characteristics. The LO1 mode is not directly observed. Additional peaks are observed at, \textit{e.g.}, 77.9, 157.5, and 196.2 mV. These can be identified as harmonics of the phonon energies, LO1+LO3, LO3+LO4, and twice LO4, respectively. The phonon energies extracted from the tunneling data are in good agreement with those observed by hyper-Raman measurements \cite{Vogt1988}, as listed in Table~\ref{tablephon}. The small difference originates presumably from the different measurement temperatures. One sample was grown using a SrTi$^{18}$O$_3$ single crystal \cite{Schneider2010}. In this oxygen--isotope substituted sample, small shifts of the phonon energies can be expected. We observed a shift of 3.2 meV towards lower energy in the LO4 mode. The other modes are shifted by less than 1 meV. The isotope effect is stronger for the LO4 mode as this mode involves large displacements of the oxygen ions. Because similar shifts have been observed in Raman measurements on SrTi$^{18}$O$_3$ (not shown), a significant fraction of $^{18}$O (larger than 60\%) has to be present in the interfacial region of the sample, even after the LaAlO$_3$ growth and annealing in a $^{16}$O environment.

\begin{figure}
\centering
\includegraphics[width=9cm]{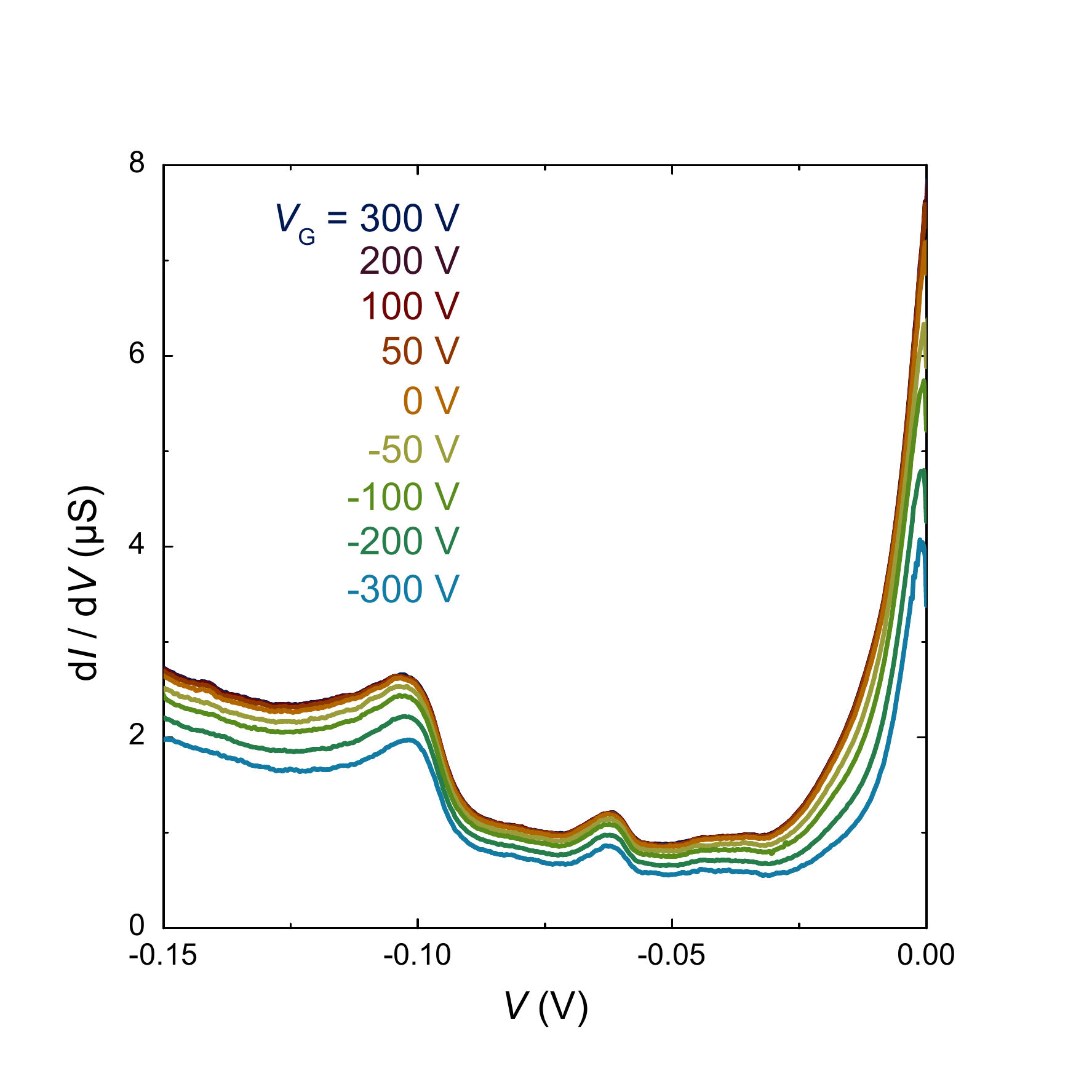}
\caption{Gate-voltage dependence of the phonon--assisted tunneling. The tunneling characteristics $\textrm{d}I/\textrm{d}V (V)$ were measured at 0.05 K. }
\label{fig3}
\end{figure}

\begin{figure}
\centering
\includegraphics[width=9cm]{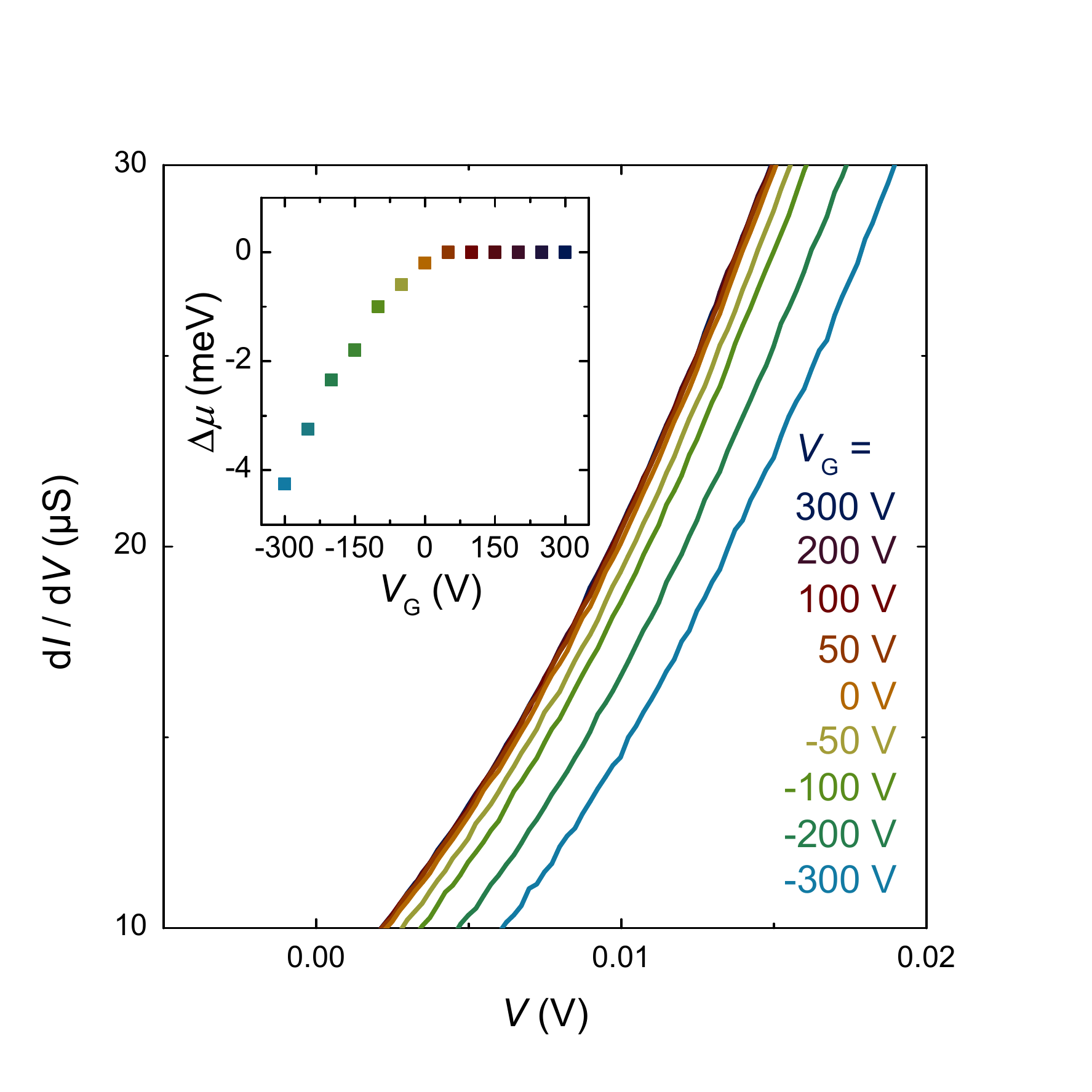}
\caption{Gate voltage-dependent conductance characteristics at positive voltages, measured at 0.05 K. The shift in voltage in these characteristics was used to determine the change in chemical potential $\Delta\mu$ that is shown in the inset. }
\label{figmu}
\end{figure}

Having shown that electron--phonon coupling can be directly observed in the LaAlO$_3$--SrTiO$_3$ tunnel junctions, we now discuss its relation to superconductivity. We measured the gate-voltage dependence of the phonon--assisted tunneling in a device for which we previously determined the superconducting phase diagram \cite{Richter2013}. A positive (negative) gate voltage accumulates carriers at  (depletes carriers from) the 2DEL. Optimum doping with maximum $T_\textrm{c}$ is achieved at $V_\textrm{G}$ = 0~V. At positive (negative) gate bias the system is overdoped (underdoped). The gate-voltage dependence of the tunnel characteristics is presented in Fig~\ref{fig3}. For positive gate voltages the characteristics are not affected by $V_\textrm{G}$. Negative gate voltages, however, decrease the tunnel conductance significantly. This decrease is due to a change in the chemical potential $\mu$ that reduces the occupied density of states in the 2DEL. To precisely determine the gate-voltage dependence of $\mu$, we analyzed the shifts in voltage in the conductance curves at positive voltages, see Fig.~\ref{figmu}. These shifts are constant over a large voltage range and can be used to accurately determine $\Delta\mu$, the change in chemical potential in comparison to the overdoped cases. We determine $\mu$ for $V_\textrm{G}$ $>$ 50 V by finding the crossing point between the almost constant conductance in the range -55 $<$ $V$ $<$ -35 mV (mostly inelastic tunneling) and the strongly voltage dependence conductance in the range -25 $<$ $V$ $<$ 0 mV (elastic tunneling). This yields $\mu$ = 30 $\pm$ 2 mV,  in good agreement with the data in Fig.~\ref{fig1}. The tunneling spectra in Fig.~\ref{fig3} show a small reduction of the conductance close to the Fermi energy at $V >$ -3 mV. In related cases, this reduction is attributed to the Altshuler--Aronov correction to the density of states of an electron system with electron--electron interactions \cite{Altshuler1979}. A detailed analysis of the Altshuler--Aronov correction is beyond the scope of this article. The most important observation is that the phonon--assisted tunneling peaks are consistently observed for all gate voltages. 

\begin{figure}
\centering
\includegraphics[width=9cm]{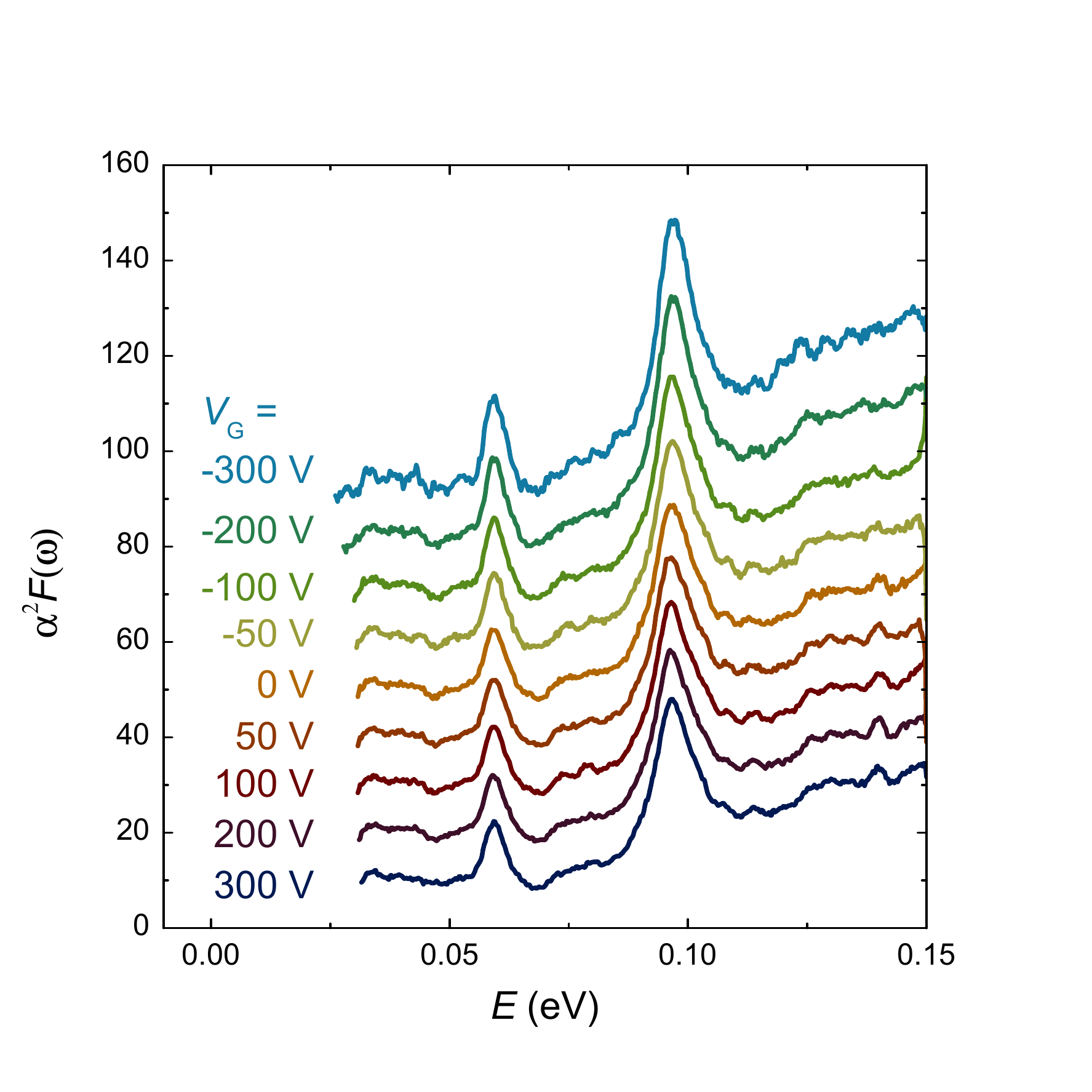}
\caption{The electron--phonon spectral function $\alpha^2F(\omega)$ as calculated from the $\textrm{d}I/\textrm{d}V (V)$ data in Fig.~\ref{fig3}. Each curve has been offset for clarity.}
\label{fig5}
\end{figure}

For quantitative analysis of the electron--phonon coupling, the inelastic tunneling probability has to be evaluated at the different energies. In the case of tunneling with the emission of real phonons studied here (as opposed to the case of the virtual phonon--coupling induced self-energy correction studied in superconducting tunnel junctions \cite{McMillan1965}), the tunneling probability is proportional to $\alpha^2F(\omega)$ \cite{Adler1971, Schackert2015}. The tunneling conductance is a function of both the density of states and the inelastic tunneling probability. The observed shape of the inelastic tunneling peaks is due to a convolution of the electron--phonon spectral function with the occupied density of states of the 2DEL. Because the voltage range in which elastic tunneling occurs ($V > -\mu$/$e$) and the voltage range in which inelastic tunneling occurs ($V < -\mu$/$e$) are separated in energy, we have experimental access to the occupied density of states of the 2DEL. Here $e$ is the electron charge. We therefore deconvoluted the density of occupied states (as measured by elastic tunneling in the voltage range -$\mu/e < V < 0$) from the inelastic tunneling conductance (see Methods). This procedure yields a function proportional to the electron--phonon spectral function, which is shown in Fig.~\ref{fig5}. The magnitude of the function has been normalized such that $\alpha^2F$\,($\omega$)$\cdot$d$\omega$ reflects the ratio of the inelastic tunneling transmission in an energy range d$\omega$ around $\omega$ and the total elastic tunnel transmission. The dominant features of $\alpha^2F$\,($\omega$) are the strong coupling at the LO3 ($\sim$60 meV) and LO4 ($\sim$100 meV) phonon modes. The phonon energies obtained from $\alpha^2F$\,($\omega$), see Table 1, are are in good agreement with those obtained from the peaks in -d$^2$$I$/d$V^2$($V$). Next to the peaks from the phonons, a background that increases with increasing energy is present. In tunneling, the barrier height decreases with increasing voltage and therefore some additional elastic tunneling is also present at voltages $V < -\mu$/$e$. The deconvolution procedure ignores this and the elastic part of the tunneling conductance results in the background. The $\alpha^2F$\,($\omega$) function is unaffected by the gate voltage, except for an overall increase at negative gate voltages. 

\section*{Discussion}
The main objective of our study is to identify the pairing mechanism of the LaAlO$_3$--SrTiO$_3$ 2DEL by measuring the coupling of the electrons to bosonic modes with inelastic tunneling spectroscopy. Up to energies of 200 meV we find only coupling to the LO phonons of SrTiO$_3$. The phonon--assisted tunneling conductance in the 2DEL junctions is significantly larger than that observed in, e.g., Pb junctions\cite{Adler1971, Dynes1975, Schackert2015} and is of similar magnitude as that observed in doped SrTiO$_3$ junctions \cite{Sroubek1969, Hayashi1981}. This indicates that the electron--phonon scattering crosssection is large. The measurements yield $\alpha^2F$\,($\omega$) with an unknown proportionality constant and we therefore cannot calculate the critical temperature. So, the results are not umambiguous proof that the LO phonons do provide the pairing mechanism. Because we observe coupling to the LO phonons and do not observe coupling to other modes, we have to conclude, however, that these tunnel spectroscopy measurements point clearly to electron--phonon coupling as the pairing channel. Note that next to the observed coupling to the LO phonons, coupling to the acoustic phonons may be present as well, as discussed for example in ref \cite{Klimin2014}. Our measurements namely do not discriminate coupling to acoustic phonons modes from the elastic tunneling conductance if the bosonic spectral function is not strongly energy-dependent. 

\begin{figure}
\centering
\includegraphics[width=9cm]{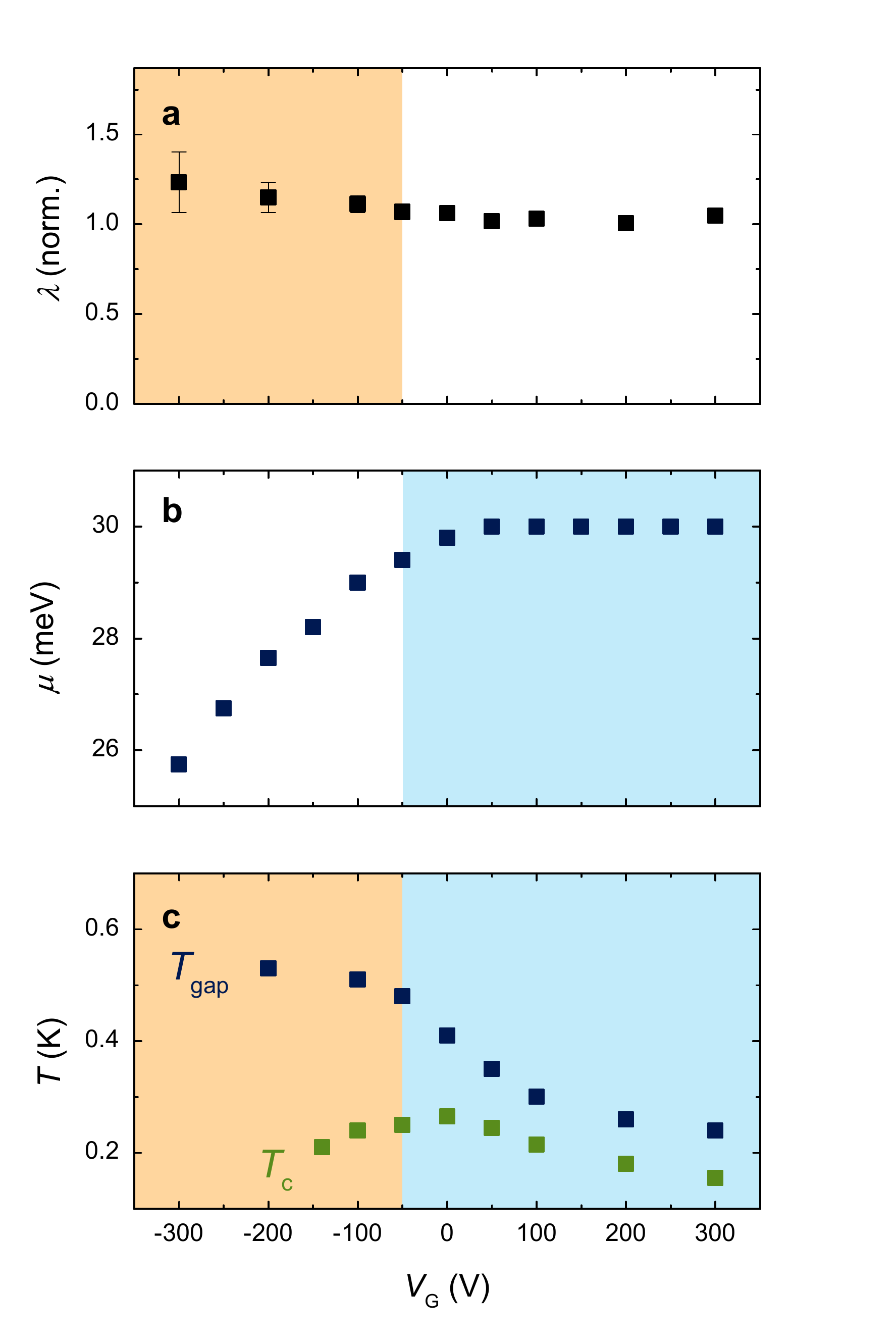}
\caption{Gate voltage dependence of a) $\lambda$, b) $\mu$ and c) $T_\textrm{gap}$ and $T_\textrm{c}$. The values for $\lambda$ are normalized to those in the overdoped region. In the underdoped region (brown background) the doping dependence of $T_\textrm{gap}$ coincides with a change in $\lambda$. In the optimally doped and overdoped regions (blue background) the reduction of $T_\textrm{gap}$ coincides with a lack of increase of $\mu$ with increasing carrier density, indicating the additional charge carriers do not reside in the band that is superconducting, but interfere. The error margins of $\lambda$ reflect the uncertainty in the background subtraction procedure. The values for $T_\textrm{gap}$ and $T_\textrm{c}$ are taken from an earlier publication \cite{Richter2013}. }
\label{fig6}
\end{figure}

We next discuss the gate-voltage dependence of the electron--phonon coupling. The coupling strength is characterized by the McMillan parameter $\lambda$ that is obtained from $\alpha^2F$\,($\omega$) by \cite{McMillan1968}
\begin{equation}
\lambda = 2 \int_0^\infty  \alpha^2F (\omega) / \omega\cdot \textrm{d}\omega .
\label{eq}
\end{equation}
We extracted $\lambda$ from the data in Fig.~\ref{fig5} by integrating equation~\ref{eq} over the energy range 30 $<$ $E$ $<$ 145 meV. The resulting values for $\lambda$ are presented in Fig.~\ref{fig6}a, where they have been normalized to those in the overdoped region. Approximately 60\% of the coupling is due to the LO4 mode and approximately 25\% of the coupling is due to the LO3 mode, independent of $V_\textrm{G}$. The doping dependence of $\lambda$ can be compared to that of the previously determined $T_\textrm{gap}$ values, the temperatures at which the superconducting gap closes. $\lambda$ is constant for $V_\textrm{G} > -50$ V and increases with decreasing carrier density in the underdoped region of the phase diagram. In the underdoped region of the superconducting phase diagram, the increase of $\lambda$ with decreasing carrier concentration is qualitatively consistent with the increase of $T_\textrm{gap}$. However, in the main part of the phase diagram $T_\textrm{gap}$ depends strongly on the applied gate voltage and $\lambda$ is constant. The decrease of $T_\textrm{gap}$ in the optimally doped and overdoped region is therefore puzzling. In the following we show that this decrease can be directly understood by considering the band structure of the 2DEL.

Density functional theory calculations \cite{Popovic2008,Son2009,Zhong2013}, transport properties \cite{Pentcheva2010, Joshua2012} and recent angle-resolved-photoemission (ARPES) data \cite{Berner2013} indicate that the 2DEL comprises several bands: small electron mass d$_{xy}$ orbital derived bands and large electron mass d$_{xz}$ and d$_{yz}$ orbital derived bands. Because the d$_{xz}$ and d$_{yz}$ orbital derived bands have larger momenta in the tunnel direction, tunneling occurs predominantly to those bands. This is consistent with the band structure determined from ARPES measurements \cite{Berner2013}: the bottom of the d$_{xz}$ and d$_{yz}$ orbital derived bands lies approximately 50 meV below the Fermi energy, in reasonable agreement with the $\mu$ = 30 mV observed in tunneling. The bottom of the  d$_{xy}$ orbital derived bands lies approximately 300 meV below the Fermi energy. Because the superconducting gap is observed in the tunneling characteristics with the temperature dependence of a primary order parameter \cite{Richter2013}, we conclude that the d$_{xz}$ and d$_{yz}$ orbital derived bands host the dominant contribution to the superconductivity. 

Figure~\ref{fig6}b presents the gate voltage dependence of the chemical potential in these bands. In the underdoped region $\mu$ steadily increases with increasing carrier density, as expected. However, at optimal doping the increase of $\mu$ levels off and $\mu$ is almost constant in the entire overdoped region. Because the shape of the tunneling characteristic is virtually independent of $V_\textrm{G}$, the additionally induced charge carriers have to reside in one or several higher-energy bands with large density of states that are not accessible to the tunneling, such as in a d$_{xy}$ orbital derived band or in a band further away from the interface. In agreement with this conclusion, signatures of such a band have been observed in transport studies \cite{Joshua2012}, exactly appearing at gate voltages larger than the one at optimum $T_\textrm{c}$ (see also the Hall effect data in references \cite{Bell2009, Herranz2015}). The charge carriers in this additional band are presumably not superconducting (or only superconducting due to the proximity effect from the other bands), but will provide more Coulomb scattering to the electrons in the superconducting bands, thereby explaining the reduction of superconductivity in the overdoped region. Electronic phase separation in superconducting and non-superconducting regions could also explain the doping independence of $\mu$, but in this scenario the density of states is expected to change with doping, contrary to the measurements.

In summary, we performed tunneling experiments to identify the superconducting pairing mechanism in the LaAlO$_3$--SrTiO$_3$ 2DEL. We determined $\alpha^2F$\,($\omega$) and observed electron--phonon coupling across the entire superconducting phase diagram. We only observed coupling to the LO phonons of SrTiO$_3$ and the coupling to the LO4 mode was measured to be particularly pronounced. We conclude that electron--phonon coupling likely provides the pairing mechanism for the superconductivity. In the underdoped region the decrease of $T_\textrm{gap}$ with increasing carrier density is possibly explained by a reduction of the electron--phonon coupling strength, but in the optimally doped and overdoped regions the electron--phonon coupling is doping-independent. In these regions an additional band becomes populated, as evidenced by tunneling measurements of the chemical potential in the 2DEL. The charge carriers in this band result in additional Coulomb scattering and thereby weaken the superconductivity, causing the reduction of $T_\textrm{gap}$. This scenario is intriguingly similar to the reduction of $T_\textrm{c}$ in overdoped gate-tuned MoS$_2$ \cite{Ye2012, Das2014} and to the constant chemical potential observed in cuprate interface superconductors \cite{Wu2013}.

\section*{Methods}

\subsection*{Experimental}
The tunnel junctions were fabricated by first growing a 4 or 5 unit cell thick layer of LaAlO$_3$ on TiO$_2$ terminated \cite{Koster1998} SrTiO$_3$ by pulsed laser deposition to create the 2DEL \cite{Thiel2006}. Then a gold top electrode was deposited on the LaAlO$_3$ \textit{in situ}. The gold layer was patterned using standard photolithography and selective chemical etching with a KI+I$_2$ solution. In a final processing step ohmic contacts to the electron system were made by argon ion milling and Ti sputtering. The fabrication and characterization of the devices is described in more detail elsewhere \cite{Richter2013}. We fabricated devices using both SrTi$^{16}$O$_3$ and SrTi$^{18}$O$_3$ \cite{Schneider2010} single crystals. We did not see an effect of the oxygen isotope exchange on the normal-state transport properties or on the superconducting properties of the 2DEL.  

\subsection*{Extracting the electron--phonon spectral function}
In superconducting tunnel junctions, the self-energy correction to the BCS density of states due to the electron--phonon coupling can be directly observed and quantitatively modeled using the Eliashberg theory. In the LaAlO$_3$--SrTiO$_3$ 2DEL this does not work because the self-energy correction is not observed. The phonon--assisted tunnel spectra are identical in the superconducting and normal state (not shown) because the superconducting gap ($\sim$50 $\mu$eV) is much smaller than the phonon energies involved. Therefore the normal state density of states has to be used to quantify the electron--phonon coupling. In normal metal phonon--assisted tunneling, the inelastic part of the tunneling conductance is a convolution of the electron--phonon spectral function and the density of states of the electron system. 
\begin{equation}
g_\textrm{i} (E) = K \int_{-\infty}^\infty \alpha^2 F (E-\tau) N_\textrm{occ} (-\tau) \mathrm{d}\tau,
\label{eq1}
\end{equation} 
where $g_\textrm{i}$ is the inelastic conductance, $K$ is a constant and $N\textrm{occ} (E)$ is the occupied density of states. In case the density of states is constant in the energy range of interest, $N\textrm{occ}$ is the heaviside step function. Then the derivative of equation~\ref{eq1} with respect to the energy yields the proportionality between d$g_\textrm{i}(E)/\textrm{d}E$ and $\alpha^2F(\omega)$. This relation is generally used in inelastic tunneling spectroscopy \cite{Adler1971, Wolfbook, Schackert2015}. In case the density of states is not constant, changes in $g_\textrm{i} (E)$ are either due to changes in $\alpha^2F(\omega)$ or to changes in $N\textrm{occ} (E)$. When $N\textrm{occ} (E)$ is known, $\alpha^2F(\omega)$  can be extracted from $g_\textrm{i} (E)$ with the procedure described in the following. A discrete version of Eq.~\ref{eq1} can be written as
\begin{equation}
\alpha^2 F (E) =  \frac{1}{\Delta E N_\textrm{occ}(0)} \left( g_\textrm{i}(E) - \sum_{\tau=\Delta E}^\mu \alpha^2 F (E-\tau) N_\textrm{occ}(-\tau) \Delta E \right).
\label{eq2}
\end{equation}   
Here $\Delta E$ is the step size in energy. The summation is cut off after $\tau$ = $\mu$ because the density of states is zero at larger energies. We identify the tunnel conductance in the range -$\mu$/e $<$ $V$ $<$ 0 as  $N_\textrm{occ}(E)$ and the tunnel conductance in the range $V$ $<$ -$\mu$/e as $g_\textrm{i} (E)$. Now $g_\textrm{i} (\mu+\Delta E)$ can be used to determine $\alpha^2 F (\mu+\Delta E)$, because $\alpha^2 F (E)$ is assumed to be zero for $E$ $<$ $\mu$. Following this, $\alpha^2 F (\mu+2\Delta E)$ can be obtained. The procedure works best when the tunnel conductance at $E$ = $\mu$ is zero, giving a clear separation between density of states and inelastic tunneling. If this is not the case, a large spike will result in $\alpha^2 F (E)$ at $E$ = $\mu+\Delta E$. This spike can be removed by adjusting a constant value for  $\alpha^2 F (E)$ for $E$ $<$ $\mu$.

\section*{Acknowledgements}
We thank A. Brinkman, P. Hirschfeld, P. Horsch, J.\,R. Kirtley, T. Kopp, F. Loder, K.\,A. Moler, N. Pavlenko, and J.-M. Triscone for valuable discussions. 

\section*{Author contributions statement}
H.B. and J.M. conceived the experiment. C.R. fabricated the tunnel junctions. C.R., H.B. and E.F.-T. performed the measurements and analyzed the data. C.S. performed the oxygen isotope exchange. J.M. supervised the research. H.B. wrote the manuscript with help from all other authors.

\section*{Additional information}

\textbf{Competing financial interests} The authors declare no competing financial interests. 

\begin{table}
\centering
\caption{The energies of the longitudinal optical phonon modes of SrTiO$_3$ as observed by hyper-Raman scattering \cite{Vogt1988}, and the present tunneling spectroscopy. The Raman data were taken at 300 K and the tunneling data were taken below 4.2 K. The numbers in parentheses indicate indirect measurements, derived from higher order processes.}

\begin{tabular}{|c| c| c|c|}

\hline
Phonon 	&Hyper-Raman      	&Tunneling 	 	 & Tunneling			\\
mode		&		&-d$^2I$/d$V^2$	&$\alpha^2 F(\omega)$\\
\hline 
LO1																		&(19.8 meV)		&(18.5 meV)	&(19.0 meV)\\
LO2																		&33.0 meV		&33.5 meV	&35.0 meV\\
LO3																		&58.8 meV		&59.4 meV	&60.4 meV \\
LO4																		&98.6 meV		&98.1 meV	&99.3 meV\\
\hline
\end{tabular}
\label{tablephon}
\end{table}
   
\end{document}